# Leverage and Uncertainty

*Mihail Turlakov[1]*

**Abstract**

Risk and uncertainty will always be a matter of experience, luck, skills, and modelling. Leverage is another concept, which is critical for the investor's decisions and results. Adaptive skills and quantitative probabilistic methods need to be used in successful management of risk, uncertainty and leverage. The author explores how uncertainty beyond risk determines consistent leverage in a simple model of the world with fat tails due to significant, not fully quantifiable and not too rare events. Among particular technical results, for the single asset fractional Kelly criterion is derived in the presence of the fat tails associated with subjective uncertainty. For the multi-asset portfolio, Kelly criterion provides an insightful perspective on Risk Parity strategies, which can be extended for the assets with fat tails.

[1] Head of CVA desk, Global Markets division, Sberbank CIB. Mihail has about 10 years' experience in financial industry in credit, FX and quantitative trading areas. He has worked on developing new business areas, modelling and trading while at RBS, Deutsche, WestLB and Sberbank. Before moving on to a financial career, Mihail worked as a researcher at the Departments of Physics in Cambridge and Oxford Universities in United Kingdom. He received his Ph.D. in theoretical condensed matter physics from University of Illinois at Urbana-Champaign in United States in 2000.

Uncertainty and risk play fundamental role in the portfolio and risk management. The more objective measurable and the more subjective non-quantitative aspects of decisions can be separated into risk and uncertainty correspondingly (Knight 1964). The risk, with known and quantifiable probabilities and outcomes, became associated with the standard deviation (or volatility) within classical expected utility theory and Black-Scholes-Merton theory. Not surprisingly, the uncertainty is associated with other less understood, yet important, concepts of unquantifiable outcomes, liquidity, partial information and subjective behavioural biases (Booth 2014). Uncertainty and risk constrain the leverage (allocation size), which would amplify the premium returns in unlimited way. While Kelly criterion (Kelly 1956) proposed optimal leverage in the presence of risk, the fascinating conceptual link between leverage and uncertainty is explored quantitatively and phenomenologically here.

*General consequences of Kelly criterion in the world with fat tails are discussed with the practical angle of the portfolio and risk management practitioner.* In the view of the complexity of modelling the processes and phenomena with fat (heavy or power-law) tails, only a simple probabilistic model is analysed quantitatively here. The paper is organised as follows. The introductory section "Risk and Uncertainty" sets the stage and discusses different aspects of risk and uncertainty. The section "Risk and Leverage" reviews the conceptual links between risk and leverage in the modern finance and introduces the Kelly criterion. The two sections "Kelly criterion in the fat tails world" and "Kelly Parity" are more technical and mathematical; these sections contain several novel quantitative results about Kelly criterion for a single asset and for multi-asset portfolio calculated in a simple toy probabilistic model of the world with fat tails. How the uncertainty, parametrised by left and right tails of the probability distribution, changes the Kelly criterion is presented in the section "Kelly criterion in the fat tails world". Basic important connections are established between Kelly Parity, based on Kelly criterion, Risk Parity, a popular risk control portfolio approach, and classic Mean-Variance Optimization in the section "Kelly Parity". The effects of the volatility skew and convexity[2] on Kelly Parity and Risk Parity are also discussed. The "Adaptive management of risk, uncertainty, and leverage" section considers practical application of statistical Kelly system in dynamic and intermittent real-world financial markets. The "Conclusion" section consolidates the general insights and presents the outlook. Simple conceptual consequences of the drawdown aversion, related to the uncertainty of the significant losses, on the efficient frontier in the classic portfolio theory are explored in the appendix "Concave Efficient Frontier". Another appendix contains the details of Kelly criterion calculations in the simple probabilistic model with a tail loss.

## Risk and Uncertainty

The key concepts of risk and uncertainty have been explored from different perspectives and play different roles in any real life decision making. The risk is about quantifiable probabilities and outcomes; the uncertainty is about degree of belief and subjectivity. From one side, classical and neo-classical theoretical economics made tremendous progress by focusing on risk in the framework of the probabilistic decision theory and the utility theory. Black-Scholes-Merton theory (Black 1973) and Mean-Variance optimisation theory (Markowitz 1959) have been developed in the rigorous mathematical formulation where the volatility represents the risk. From another side, the practical finance and risk-management practices proceeded by realising *the duality[3] between risk and uncertainty* and defining the VaR boundary separating them (Brown 2012). The inefficiency of the markets (Shiller 2005) and the behavioural biases and aversions (Kahneman 1979) have been also recognized in driving the uncertainty of the markets. Markets are uncertain as much as risky.

---

[2] The volatility skew and convexity mean linear and quadratic adjustments as a function of the underlying asset. Equivalently, the (sufficiently large) volatility skew and volatility convexity (smile) in the Black-Scholes based description mean asymmetric left and right fat tails of the probability density function.

The uncertainty has many subjective and non-quantifiable, in the sense of Keynes and Knight, aspects. The uncertainty can be associated with fat tails of the probability distributions. Although this is certainly an oversimplification, I will make the following proposition; *the risk and the uncertainty can be associated with the centre and the fat tails of the probability density functions (pdf's) correspondingly*. In this paper, I use this simple proposition, which allows advancing the understanding of the uncertainty. Formally, under a unit measure, (the total probability is one), the tails is what remains after the possible risks are quantified. In each present moment, the price is clearly certain within bid and offer prices for any, unless very illiquid, asset. The markets are more certain about the centre of the distribution, while the investors differ in their subjective views more strongly about the tail outcomes[4]. To expand on the above proposition, subjective probability distribution functions can be thought to be incorporating all those subjective factors (limited information, behavioural biases, liquidity information, limited capital, etc.) into the tails of the distribution functions of the individual investors. Yet the centre of the distributions is quite constrained for all investors by the current market valuations and "current markets consensus". Subjective view on the risk can definitely be different from the collective market view, but collective markets efficiency imposes strong constraints at not too long time horizons. At longer time horizons and larger deviations[5], the uncertainty becomes more important, and notwithstanding the markets efficiency, the markets are much more time dependent and inconsistent than the current pricing (of the forward curves and the options at different maturities) suggests. Essentially, by their construction and their actions, the markets explore and "quantify" the central "phase space"[6] of possible risks[7], and the remote parts of the phase space represent the uncertainty. These remote parts of the pdf, which have lower probability and longer time[8] to be reached, are naturally assigned to the tails.

Several other aspects of the uncertainty are notable. Firstly, the subjective difference between the price and the value is another definition of the uncertainty. At the height of panic or irrational exuberance, the uncertainty as well as the mispricing are at the highest. Risk perception may differ from risk reality, and the uncertainty is in the eye of the beholder. Secondly, the uncertainty has a systemic component, due to the collective financial system, and the idiosyncratic components due to unpredictable events (political, social, technological, etc.). Thirdly, although many possible unpredictable events and their unstable sequences may occur, all these uncertainties need to be codified and lumped together into the fat tails. In other words, the markets are clearly incomplete in the tails.

---

[3] interconnectedness and complementarity
[4] In particular, at-the-money and shorter-dated options are more liquid, while out-of-the money and/or longer-dated options are less liquid and have wider bid-offer spreads.
[5] The delineating time horizon between the risk and the uncertainty seem to depend on the type of uncertainty itself.
[6] The phase space (conventionally used in physics and other sciences) means the full space of all relevant variables of the system.
[7] The fluctuations of the markets, due to news flow and all other factors, test the views and the positions of the investors; this process reveals the balance of "local risks". Perhaps, in other words, the risk (volatility) are "local risks", and the uncertainty is "remote risks".
[8] The property of markets to trend signifies the adaptability and confluence of various market mechanisms and participants (capital flows, leveraging, regulatory pressures, typical behaviour, etc.). Markets is a complex and non-ergodic system beyond simple physical systems, which explore the phase space progressively and ergodically.

In reality, risk and uncertainty mix and not clearly delineated, while they might only be separated in a graphical representation of pdf (probability density function) or any kind of a model. The modelling of uncertainty is paradoxical, and the models are only metaphors, which should not be taken too seriously (Derman 2012). After all, markets would not be needed if a perfect (deterministic or stochastic) model existed, and the markets is efficient mechanism, which is as of now "the worst, except for all others". The related phenomena of subjectivity, diversity and uncertainty appear to be essential and intrinsic for the competitive markets. Several general points illustrate further the dichotomy and complementarity of risk and uncertainty. Firstly, the diversification, one of the pillars of modern finance theory, is proposed to reduce the 'risks'[9], yet the financial 'risks' do not disappear, but they transform into different shapes and forms. To put it differently, *the diversification rebalances the 'risks' between the "familiar risk" around the centre of the distribution (the volatility) and the "uncertain risk" of the tails of the distribution (the uncertainty)*. Secondly, a single realization of the future possibilities carries enormous importance for the investors behaviourally, sociologically (Nelson 2014) and organizationally (in the sense of agent-agency problem). The history is a single path, not "Monte-Carlo averaged"[10]. The finite sample of important investor's decisions is ever present in mixing risk and uncertainty[11], although the expected utility theory does not distinguish between a single and multiple bets decisions (Samuelson 1963). Thirdly, in the absence of the model[12], the centre and the tails cannot be separated unambiguously, since the volatility of the markets is time-dependent. For example, "Bayesian adjustment" in GARCH models corrects what would be a naïve mixing of uncertainty and risk (Arnott 2005).

*The mathematical description of the various real-life risks and the uncertainty is hard and illusive*, and in any case the high level of skill and experience is required in the application of the quantitative models successfully in the real financial markets. It seems fair to say that, although the empirical presence of fat tails in financial time series is widely accepted, there is no well-established and accepted dynamic theory in finance[13], which describes realistically the fat tails. Major advances ( (Mandelbrot 1997), (Sornette 2004), (Engle 1995), (Mantegna 2000)) have been made to describe fat-tailed distributions, but there are principal and conceptual differences between the markets and natural sciences. The limitations are non-repeatability of history and the adaptability of the financial markets to name a few. Modelling of outlier events can be problematic, as a matter of principle, and the models have to be used sceptically. The quantitative models tend to be recalibrated on the new market data and lack the predictive power of the models in the natural sciences. Yet, even in the absence of scientifically sound dynamic models, the fat tails are the definitive manifestations of the transitive dynamic reality and behavioural tendencies of the markets, and these innate properties must be considered in any portfolio management quantitatively and qualitatively, just like in the modern options pricing and management.

---

[9] For no better word, 'risks' are used here and below only in the very general sense of the word. 'Risks' mean both the risk and the uncertainty of (Knight 1964) together without differentiating between them.

[10] In finance, many future expectations are calculated by averaging over multiple future outcomes generated by Monte-Carlo method.

[11] Quoting from "Risk and Uncertainty" of (Samuelson 1963): "… there will necessarily be left an epsilon of uncertainty even in so-called risk situations. As Gertrude Stein never said: Epsilon ain't zero. This virtual remark has great importance for the attempt to create a difference of kind between risk and uncertainty in the economics of investment and decision making."

[12] For an important theoretical example, in the mathematical language, the binomial and Gaussian models possess only risk and none of uncertainty.

[13] Some models in physics are phenomenological, but, arguably, the most powerful theories in physics are necessarily dynamic.

The fat-tailed nature of the distributions in finance is all-important and inherent, rather than minor or secondary, property of the markets (Mandelbrot and Taleb 2010). "What risk really means" is the very important question, and I think that uncertainty and possibility of a loss associated with the downside fat tails is the most relevant threat. "What they fear is the possibility of permanent loss" (Marks 2015). Thus, the risk management of the uncertainty and fat tails is of critical importance. A recent model-free analysis of historical data showed that for majority of the trading strategies (with several specific exceptions) *the risk premium is largely due to the presence of the skew or equivalently downside fat tail* (Lemperiere 2015). This major empirical finding[14] incentivises deeper research into the connections between the uncertainty (fat tails) and the leverage (capital allocation), which is the topic of this paper.

## Leverage and Risk

The leverage - the allocation size - is as important as the odds (the probabilities) of the outcomes for the capital allocation decisions. The leverage needs to be decided in *absolute* terms with respect to the investor's available capital and in *relative* terms with respect to relative allocation between different assets (or trading strategies) within the investor's portfolio. In the modern portfolio theory (Markowitz 1959), the relative allocation between different assets, based on their expected returns, volatilities and correlations, is the focus of the theory. The leverage financing of the modern financial system permits an investor to choose the absolute leverage among other risk preferences. Leverage aversion changes the predictions of the portfolio theory (Assness 2012), and modern Risk Parity (RP) strategies exploit this opportunity. Leverage is a mechanism and a tool, not a danger in itself, while uncertainty and permanent loss can definitely be associated with danger.

Absolute leverage is a risk decision determined by internal and external (due to the perception of the counterparties providing leverage financing) factors. Uncertainty, rather than risk, appears to be a more determining factor for the absolute leverage, since both, absolute leverage and uncertainty, are more collectively determined properties with individual variations. Qualitatively, *uncertainty appears to be more important determinant of the absolute leverage, risk determines relative leverages*. The collective effects of aggregate market liquidity and leverage have been elucidated (Adrian 2010) by pointing out the role of financial intermediaries and credit cycles. Liquidity and uncertainty are complex and different for different market participants. Market liquidity, funding liquidity (Brunnermeier 2008), absolute and relative leverage, and uncertainty are intimately linked by direct and indirect mechanisms, and are very much parts of the markets dynamics and challenges. Overall, the uncertainty and the liquidity shed light on each other[15]. Both concepts are hard to define fully even qualitatively, and both are complex and multi-dimensional[16].

After the above overview of general aspects of the leverage, I would like to introduce the mathematical Kelly criterion (Kelly 1956), which prescribes *the optimal allocation of the current capital* for the 'best long run'. Suppose a favourable outcome wins *b* units for every unit wager. Further, suppose that for each trial the win probability is $p > 0$, the lose probability is $q$, and the expectation of the game $pb-q >0$ is positive. From optimization of the geometric growth rate, the optimal fraction of the current capital is $f^* = (bp - q)/b$. In the continuous limit of the random Gaussian process with the expected return µ and the volatility $\sigma$, Kelly allocation is a dimensionless fraction [17] of the available capital $f^* = \mu/\sigma^2$. The optimal geometric growth rate is equal to $g^* = \frac{1}{2}\mu f^*$. The quantitative and statistical power of Kelly system rests on accurate estimate of the risk parameters and multiple bets "averaging".

---

[14] The skew nature of risk premium appears to be more naturally sustainable by the markets unlike axiomatic risk premium, defined as volatility, of classical Capital Asset Pricing Model (CAPM).

[15] For instance, low uncertainty suggests using higher leverage, and, just as well, high liquidity permits higher leverage in the trading.

[16] Yet the past risk, in the sense of the volatility, can be calculated as the historical standard deviation of the past time series as well as future risk (future volatility) is traded and forecatsed by the options markets.

In the Gaussian (and binomial) model, in the absence of uncertainty (tails), the optimal leverage is fully determined by the risk (the volatility) in the optimization problem for a given return. For the binomial outcomes, the optimal allocation has several nice properties (Thorp 2006); (a) maximizing the geometric growth rate, (b) minimizing the probability of ruin, (c) the penalty in reduced growth rate for moderate underbetting, which offers a margin of safety, is not severe. The drawbacks of Kelly system are as follows: (a) the intermediate returns are volatile roller coasters (due to a constant proportion betting), (b) the drawdowns can be large, and (c) ruin is possible. The practical application has also several challenges (M. J. Mauboussin 2006): (a) the performance of money managers is frequently evaluated on arithmetic returns, not geometric returns essential in Kelly system, (b) possible large drawdowns and loss aversion can bias investors away from Kelly system.

## Kelly criterion in the fat tails world

Kelly criterion is a major theoretical insight, which provides a practical guideline, but the application requires certainly adept skills and experience (Thorp 2006). In order to make mathematical progress to extend Kelly criterion to the world with the uncertainty, we will have to parametrise the tails of the pdf. Due to the theories of Savage and others, subjective probabilities can be reduced to mathematical probabilities, although not always fully consistently (Ellsberg 1961). The section "Adaptive management of risk, uncertainty, and leverage" will discuss the practical aspects of the application of the extended Kelly criterion, given the conceptual differences between uncertainty and quantifiable risk, in the real world.

Our aim in this section is to develop *Kelly criterion in risky and uncertain world*[18] in the presence of fat tails. The main quantity of the interest is the total wealth (capital) after many (N) rounds of the financial bets $W = \prod_{i=1}^{N}(1 + f_i X_i)$, where $f_i$ is the fraction of the capital invested in the round $i$, and the (negative or positive) outcome is $X_i$. The problem is to find the optimal fractions $f_i$ in order to maximize the final wealth and to minimize the unfavourable outcomes and risks.

In the presence of the tail losses of the probability $\alpha$ and the size $ETL$ and tail wins of the probability $\beta$ and the size $ETW$, the generalized results for the leverage $f_1^*$ and the growth rate $g_1^*$ of a single asset in a simple toy model are

$$f_1^* \approx \frac{\mu}{\sigma^2} \frac{1 + \frac{\beta \cdot ETW - \alpha \cdot ETL}{\mu(1-\alpha-\beta)}}{1 + \frac{\beta \cdot ETW^2 + \alpha \cdot ETL^2}{\sigma^2(1-\alpha-\beta)}}$$

$$g_1^* \approx 0.5(1-\alpha-\beta)\frac{\mu^2}{\sigma^2} \frac{\left(1 + \frac{\beta \cdot ETW - \alpha \cdot ETL}{\mu(1-\alpha-\beta)}\right)^2}{1 + \frac{\beta \cdot ETW^2 + \alpha \cdot ETL^2}{\sigma^2(1-\alpha-\beta)}}$$

---

[17] Kelly fraction $\mu/\sigma^2$ is a more natural and dimensionless risk-adjusted measure than Sharpe ratio $\mu/\sigma$, which depends on the time scale. In other words, the risk is the variance, more appropriately than the volatility (where the risk factor is being proportional to the volatility in CAPM). See also (Arnott 2005).
[18] 'Risky world' means the world with the risk (the volatility), while 'risky and uncertain world' means the world of the volatility and fat tails. These worlds can be also called "Mediocristan" and "Extremistan" (Thorp, Mizusawa 2013).

The constant fraction betting is assumed, and the calculations are given in Appendix A. In the case of $\beta \cdot ETW = \alpha \cdot ETL$, the volatility convexity, which results in the symmetric fat tails, decreases the growth factor in two ways. First way is the multiplier $(1-\alpha-\beta)$, which is a mere consequence of the probability weights taken away from the centre to the tails. Second way is the multiplier $1/(1+\frac{\beta \cdot ETW^2 + \alpha \cdot ETL^2}{\sigma^2(1-\alpha-\beta)})$, due to larger variance associated with the tails. The remaining effect additional to the convexity effect is the skew effect (the volatility skew) on the growth factor due to the multiplier $\left(1 + \frac{\beta \cdot ETW - \alpha \cdot ETL}{\mu(1-\alpha-\beta)}\right)^2$. For the simplicity, in the absence of the right tail the skew effect of the fatter loss tail shows up as a correction $(1 - \frac{2\alpha ETL}{\mu(1-\alpha)})$ in the leading order of the $ETL$ parameter. Thus, the convexity and the downside ($ETL$ related) skew reduce simple Kelly fraction $f_0^*$ and the growth rate $g_0^*$. These results[19] provide a more detailed understanding why fractional Kelly ratio $f_1^*$ must be used in practice (Thorp 2006). The important difference from Kelly criterion in risky world is that the tail parameters $\alpha, ETL, \beta, ETW$ are subjective, and therefore the expectations about the growth rate $g_1^*$ can differ widely among different investors.

The uncertainty means that no model can express the parameters of the tails accurately; nevertheless, the above results show that there are several qualitative fractional effects due to the weight shift $\alpha$ to the tails and due to the tail loss $ETL$. All effects are significant (of order of fraction of unity), while the tail impact parameter $\frac{2\alpha ETL}{\mu(1-\alpha)}$ can easily make the growth rate negative and therefore uncertain for the realistic parameters[20]. The chosen pdf function is a toy model, which nevertheless gives important qualitative insights, while the realistic power-law pdf would have to be investigated numerically. Although an asset may or may not have a sufficiently skewed right tail for high positive capital growth, the non-linear payoffs of the derivatives can have a right-tail skew. (Thorp, Mizusawa 2013) investigated in detail the capital growth of the vanilla call options aiming to maximize the geometric returns with limited drawdowns[21]. In the world of fat tails ("Extremistan"), they found that options portfolio attains better regions of the geometric growth frontier than the underlying asset and have still better profile of the drawdowns.

---

[19] It is also clear that the leverage and growth rate can have so called convexity adjustments due the volatility of the volatility.
[20] For instance, $\alpha = 20\%, ETL = 30\%$ and $\mu = 6\%$.
[21] The wealth is a random multiplicative process, which is dominated by the most probable value in the finite sample rather than its "true average value" (Redner 1990). The crossover to the asymptotic "true average" growth of the product of N independent random numbers in the sampling is exponentially slow in N, and the impact of tail loss on the crossover and on the dominance of Kelly strategy are challenging issues. The path-dependent results should surely depend on the loss size ETL, and this is why the management of the drawdowns (limiting the losses) is critical.

The intuitive and tutorial understanding of the power of the geometric growth and the pitfalls of the additional extreme outcomes is illustrated by a simple numerical example[22]. A run of 100 even bets (1 and -1 outcomes) with 60 winning bets and 40 losing bets is offered. Kelly prescribes betting each time 20% of the current capital, which is the winning edge difference between winning and losing. After 100 bankrolls, the optimal betting will give the "average geometric return" of 749%. For a realistic sequence of only 100 bets, several "extreme sub-sequences" will dominate; see below for the discussion of the statistics of the multiplicative processes. Now suppose large losses or wins can happen. We can calculate the growth rate in three more cases; Tail Loss (TL) for one bet out of 100 of losing three times your bet, Tail Win (TW) for one bet out of 100 of winning three times your bet, Both Tails (BT) for one bet each of winning and losing 3 times your bet. The growth rates are correspondingly $g_{TL} = 366\%$, $g_{TW} = 1246\%$ and $g_{BT} = 545\%$. The variations of the growth rates are large, because the optimal Kelly bet is highly sensitive to large tail wins and losses as well as to numerical effects when the parameters of original losses/wins and tail losses/wins are comparable.

Kelly system provided key guidelines (M. J. Mauboussin 2006) for the practitioners in spite of some objections from the mainstream economists. The mainstream economics focused on developing probabilistic ensemble-average utility theory. Recent breakthrough understanding (Peters 2011) of the differences between ensemble-average and time-average formulations presents Kelly criterion in a new light and proposes the dynamic approach to the gamble problem. The non-ergodic multiplicative processes can simulate the market dynamics and the fat-tailed distributions, and the time-average formulation might provide better insights into the dynamic leverage strategies beyond the optimization calculus of Kelly criterion[23].

## Kelly Parity

In this section, I discuss how Kelly criterion applies to the multi-asset portfolio in the world with or without fat tails. Important relations and differences between different portfolio approaches of Mean-Variance (MV) optimization, based on classical MPT, Risk Parity (RP) and Kelly Parity (KP) (to be defined below) emerge. Risk Parity is a risk-adjusted portfolio management approach, which goes beyond the leverage aversion of conventional MPT (Assness 2012). The most naïve RP approach is to allocate each asset in the Risk Parity portfolio according to the risk-adjusted weight $1/\sigma_i$, inversely proportionally to the asset's volatility $\sigma_i$. A portfolio approach, which uses Kelly criterion for its construction, can be called Kelly Parity. At the level of the definitions, Risk Parity pares assets based on their risks (usually understood as the volatilities), while Kelly Parity pares assets based on their growth rates.

Kelly criterion in risky world gives the following allocation $f_i$ for the asset $i$ of the N-asset portfolio (Thorp 2006)

$$f_i = \sum_{j=1}^{N} C_{ij}^{-1} M_j,$$

---

[22] See pages 175-176 of (Brown 2012). I modified slightly the example, since the case of initial 20% probability edge combined with *one-sided* possibility of 5-fold gains leads to the growth divergence.

[23] In our analysis of Kelly criterion throughout the technical part of the paper, we assume that the risk premium and the fat tails are unrelated, in the future research it will be interesting to examine the risk premium being due to the fat tails in the context of the recent empirical finding (Lemperiere 2015)

where $C_{ij}^{-1}$ is the inverse matrix of the covariance matrix $C_{ij} = \rho_{i,j}\sigma_i\sigma_j$, and $M_j$ is the risk premium vector. Kelly Parity prescribes allocations for all assets with the total portfolio leverage $Z = \bar{1} * C^{-1} * M$. In practice, there are constraints on specific and total leverage[24], and therefore a finite selection of the assets has to be made.

In the framework of MPT and MV optimization, the maximization of Sharpe Ratio = $f^T M/(f^T * C * f)^{1/2}$, subject to the constraint $f^T * \bar{1} = 1$, has to be done. Remarkably, the solution (for instance, see (Chan 2008)) is equivalent to the Kelly solution with the absolute leverage equal to one. This simple, not widely known, result is worth noting. Namely, *Kelly Parity prescribes not only the optimal absolute portfolio leverage but also relative leverages of the assets identically to the mean-variance optimization* (Markowitz 1959). Kelly Parity[25] not only maximizes the long-term growth but also minimizes the variance for a given return (by the virtue of the Sharpe ratio minimization).

Risk Parity and Mean Variance portfolio approaches are generally viewed to be different by the construction. Risk Parity optimizes the variance without any reference to the expected risk premiums, which are generally less predictable than the average volatilities, while Mean Variance optimizes the Sharpe ratio in the space of the expected returns and the variances. It appears that RP has "extreme parameter ignorance" in the sense of ignoring the input of the expected returns[26]. Kelly Parity suggests a new perspective and a deeper connection between RP and MV. Without going into the dependencies on correlations[27], and, for the simplicity in the limit of zero correlations, *Kelly Parity is equivalent to Risk Parity if all assets have the same Sharpe ratios*. Kelly allocation $f_i = \mu_i/\sigma_i^2$ becomes equivalent to the simple $f_i = SR/\sigma_i$, where Sharpe ratio factor is the same for all assets $SR = \mu_i/\sigma_i$. Sharpe ratio factor is the overall leverage scaling remaining in RP approach to target particular return level. In fact, it can be argued that RP makes implicit assumption of assets having equal Sharpe ratios in the efficient markets, since from the beginning the focus of RP is on the balancing the assets by their corresponding risks[28].

Risk Parity encompasses a whole family of the strategies and implementations by building on the sound idea of risk-adjusted allocation. Tail Risk Parity (Alankar 2012) is a radical rethinking of RP, when the risk is defined as Expected Tail Loss, similar to the tail loss $ETL$ discussed above. TRP aims to control the tail risks without over-constraining the local variance risks. In contrast, Mean-Variance optimization minimizes variance for a given level of return[29], but the minimization of the variance leads to increasing the large risks (fattening the tails) (Sornette, Andersen and Simonetti 2000). Kelly growth calculations are insightful and practical because the risk premium and tail risks can be treated in a balanced way.

---

[24] Specific expressions for the one and two assets cases are given here. In the case of one asset, $Z = \mu/\sigma^2$ is exactly Kelly leverage. In the case of the two assets, for instance, the first asset has absolute leverage $Z_1 = (\mu_1/\sigma_1^2 - \rho\,\mu_2/(\sigma_1\sigma_2))/(1-\rho^2)$, relative leverage is $Z_1/Z$, where the total leverage is $Z = (\mu_1/\sigma_1^2 + \mu_2/\sigma_2^2 - \rho(\mu_1 + \mu_2)/(\sigma_1\sigma_2))/(1-\rho^2)$.

[25] in risky world but not necessarily in risky and uncertain world
[26] In other words, RP is agnostic about the expected returns.
[27] This technical point can be investigated separately in the future.
[28] We will not delve here into the practical application of Kelly Parity and Risk Parity based on the assets or factors (Bhansali V. 2012) and practical successes of RP due to secular trends and structural reasons.
[29] In some constructions, Risk Parity determines the relative allocation of the assets in the portfolio by leveraging them to the same return.

The close connections between all three different portfolio approaches (MV, RP and KP) show that Kelly Parity is on the solid ground for the portfolio construction 'in the risky world'. More challenging to investigate next is how Kelly Parity can be extended for skewed and convex assets. In what follows, we generalize how individual skews and convexities of each asset add up or balance each other in *Kelly Parity of multi-asset portfolio in risky and uncertain world*.

Our next task is to elucidate qualitative aspects of the skew and convexity effects at the simple conceptual level of Risk Parity and Kelly Parity portfolio approaches[30], while the technical mathematical analysis for the multi-asset portfolio with power-law tailed assets have been attempted before (for instance, (Bouchard, et al. 1998)). In the simplest case if assets are uncorrelated, the rescaling of the leverage is independent. Nevertheless some assets, previously selected to the most optimal portfolio, can become less attractive and upended by other assets due to the different suppression of the growth rates by downside fat tails. More generally, we can consider two correlated or anti-correlated assets. The two assets can have the skews co-aligned or anti-aligned with the assets correlation. For instance, skew co-alignment means that the implied and realized volatilities increase for both assets in their typical correlated move. In terms of the toy model explored in this paper, skew co-alignment means that both assets are correlated not only occurring in their normal volatility states simultaneously but also in their corresponding tail loss states. The full technical details will need to be presented in a separate future paper, and the portfolio solutions can depend in complex ways on parameters. Analytical results can be viewed as tail loss perturbation corrections to Kelly Parity portfolios in terms of small parameters $\frac{\alpha\,ETL}{\mu}$, $\frac{\alpha\,ETL^2}{\sigma^2}$ and $\frac{\mu\,ETL}{\sigma^2}$. Numerical results can also be calculated for the realistic probability density functions. In simple cases, natural results can be expected with further suppression of the growth rate in the case correlated co-aligned skews; for instance, the factor $\alpha \cdot ETL_1$ is replaced by $\alpha \cdot ETL_1(1 + \mu_2 ETL_2/\sigma_2^2)$. More interesting case is anti-correlated assets with skew alignment, in this case the absolute leverage and growth rate may not be suppressed noticeably. How a multi-asset portfolio can be optimised relative to (several) possible extreme scenarios is a separate interesting problem, which is similar to Tail Risk Parity (Alankar 2012). Multi-asset portfolio choice and optimization can aim to balance simultaneously the diversification in the few scenarios of the tail events and the amplification of the growth rate at the normal volatility core.

## Adaptive management of risk, uncertainty and leverage

Active Risk Management (RM) can embrace the complex interplay between risk, uncertainty, and leverage guided by the quantitative principles. Human decision making with its behavioural aspects need to be brought together with the quantitative insights in the overall RM framework. After all, human decisions and adaptability remain central to the financial markets and portfolio management, at least at the current, still quite basic, stage of the artificial intelligence and the quantitative understanding of the complex systems, markets and economics.

Fear, greed and some rationality drive the financial markets and the financial decision-making. In the prospect theory, (Kahneman 1979) underlined the choice value "… assigned to gains and losses rather than to final assets and in which probabilities are replaced by decision weights." The certainty is preferred to the uncertainty; moreover, the individuals tend to be risk averse in the face of gains and risk seeking in the face of losses. The uncertainty of fear dominates the uncertainty of greed; this asymmetry sustains a negative skew of equity markets. All markets have symmetric or asymmetric fat tails expressed in the volatility skew and convexity. Furthermore, Ellsberg paradox (Ellsberg 1961) illustrates how the uncertainty and the complexity reduce the risk appetite. The financial decision-making has to recognise how the behavioural views on the uncertainty feeds into the quantitative choice of leverage.

---

[30] Another interesting direction of the research is how the trend strategy can be combined with Risk Parity and Kelly Parity and hedge their skews and convexity. The trend strategy is positive convexity (Bouchard, et al. 2016) and positive skew, while Kelly Parity and Risk Parity in their simplest forms are naturally negative skew and positive convexity.

The drawdown aversion can express the fear of the permanent loss. The drawdown, defined as a certain percentage loss of the capital, is a natural and important risk measure from the prospect theory perspective, and the associated aversion to the drawdown has to be managed accordingly with behavioural individual preferences. The conventional MPT and CAPM, dealing with the final assets, as per above quote, can only consider the drawdown aversion as an afterthought (see the Appendix B). How the drawdown aversion, expressed through the uncertainty of the tail loss, changes the single-asset allocation (Kelly leverage) and the multi-asset portfolio theory, is one of the central topics of this paper. The drawdown aversion is rational, quantified by fractional Kelly criterion for fat-tailed assets, and behavioural, and both aspects are self-reinforcing.

In the absence of the tail risks, the leverage has to be scaled inversely with the variance (if the risk premium remains fixed) during stable periods of the variance. Kelly system underlined practical risk-management systems of several, if not many, successful practitioners. "The natural risk" is the variance (Arnott 2005) (Kelly 1956), not the volatility. Kelly fraction is an important dimensionless risk measure, which is different from the Sharpe ratio. "Thus the Kelly investor will dynamically reallocate as optimal $f^*$ changes over time because of fluctuations in the forecast…" (Thorp 2006) of the excess return and the volatility[31]. Even in the risky world, the challenges are many from quantitative to behavioural. Among quantitative ones are the evaluation of the risky parameters and the edge of the Kelly system showing up only slowly in the number of the financial bets in the crossover from the most probable to the "true average" (Redner 1990). The numerous behavioural ones (Peterson 2007) are the rational view on the possible long drawdowns, not "taking profits too soon", and other "path dependencies". In the world with the uncertainty tails, Kelly criterion is one of the guidelines for making decisions. The range of the decisions can be narrowed, but not to an obviously optimal choice, since the uncertainty is much less certain than the risk[32]. This is always the case in the real world as Kelly criterion can rarely be used literally.

*Risk taking and risk management are the art and the skills of making good and certain decisions under uncertainty*. The perceived uncertainty[33] and the chosen leverage are the parts of the various positive and negative feedback loops, which enforce the adaptation of the collective markets and the individual investors to new information and new strategies. The interaction of adaptive investor and adaptive markets[34] is the dynamic capital allocation game.

The risk management regimes can be classified into the three regions by the magnitude of the market downside fluctuations, i. e. relative losses (consistently with the prospect theory). The normal volatility region ("white swans") is around the centre of pdf within -10%. The transitional area between -10% and -25% can be called "grey swans", the precursors to "black swans". Lastly, the "black swans" (Taleb 2008) region of the tail beyond -25% moves. Each region requires different operational modes of the decisions and actions (Bhansali 2014) in terms of active portfolio management. Diversification and dynamic rebalancing are appropriate risk-management approaches in the normal volatility region. The transitional region is of the pivotal importance[35], connecting the normal region with the fully unpredictable region of "black swans", and this is where the uncertainties and risks are most finely balanced. The experience, the transitional hedging strategies (the options and momentum strategies are particularly pertinent), and the adaptive rational decisions can make huge difference. "Grey swans" is the key region, where the risk manager's behaviour, the risk judgement and the quantitative framework become complementary. Since the "black swan" region is extremely rare and defined as completely unpredictable, then the best way to think about "black swans" is to take considerate care of "grey swans".

---

[31] The fluctuations of the volatility (volatility of volatility) and the fluctuations of the tail parameters (tail probability and tail loss) can produce so-called convexity adjustments.
[32] However obvious or tautological this statement is.
[33] The evaluation of the uncertainty and the risks is one of the main behavioural and rational tasks.
[34] The important problem is how single-agent strategy is consistent with the multi-agent system representing the markets. A consistency of multi-agent beliefs and risk views with their interactions is far beyond this paper.

*Risk-management system is a behavioural and quantitative framework, which embodies "a talent of managing events as they arise".* The risk-management system has to protect the competitive edge[36] from deep or unrecoverable setbacks. One of the main consequences of the uncertainty, made clear by Kelly calculations, is very familiar; *do not underestimate the uncertainty, do not overleverage.* The hedging of tail risk (Bhansali 2014) is a necessity not a luxury. Even diversified portfolio has some systemic and concentration risks from the threat of some particular events. Different assets and hedges have not only different instrinsic (relative to capital structure, i.e. equity or debt) but also payoff and instrument based (options, CDS, etc.) convexity and skew, and these non-linear effects are critical in the management of grey swans. The portfolio of different financial assets (equity, debt, etc.) and instruments (OTC derivatives, futures, ETFs, etc.) have to be scrutinized for sufficient market and funding liquidity when liquidity can be hard to obtain. The uncertainty of dangers and opportunities (Brown 2012) requires proxy hedging and imaginative approaches. Tail hedging is an asset-allocation decision, and different mentality, organisational setup and behavioural biases of managing the core portfolio versus the tail risk hedging can benefit the overall portfolio.

*"Edge is the key" is the main principle of risk taking.* In the world of the quantifiable risk, Kelly criterion builds on the amplification of the risk premium (edge) through the leverage. Maximizing the long-term multi-period[37] growth rate and minimizing the ruin chances is exactly how Kelly criterion strikes the fine balance and makes the edge effective. In the great debate about the advantages and disadvantages of the active and passive management, the skilful management of leverage and uncertainty is where the active management can lead by exploiting the leverage and the professional judgement about uncertainty.

Due to the intermittent nature of the markets, successful management of leverage and uncertainty have to be path-dependent and adaptive. The origin of risk premiums is critical for the appropriate investor strategy, although the risk premium (the edge) is assumed a priori in the Kelly calculations. Different cases of risk premiums and uncertainty require specialised firms and different adaptive strategies[38]. The basic balance of greed and fear can be seen already from simple expressions of the toy model for the arithmetic growth factor $r = (1 - \alpha)\mu - \alpha \cdot ETL$ in the presence of however small probability $\alpha$ of a tail loss. Although the greed of the risk premium $(1 - \alpha)\mu$ can exceed the expected losses of the fear $\alpha \cdot ETL$, a onetime loss $ETL$ can be significant. The investors attempt to outsmart each other by banking on greed gains and trying to avoid the fear losses. Important sequence of the good outcomes in the finite sample (relative to market odds in the probabilistic sense) by luck and skill is not underestimated by many investors in comparison with the statistical long-run edge, and this observation creates important dynamics of the markets. The markets do permit relatively stable risk premiums and volatility regimes over some time periods, which can be exploited by statistical Thorp-Kelly systems, but regime changes have to be recognized by the outside adaptable decision maker.

---

[35] In the note of caution from the famous investor, Peter Lynch: "Far more money has been lost by investors preparing for corrections, or trying to anticipate corrections, than has been lost in corrections themselves".
[36] In the Kelly calculations in this paper, the edge is assumed to be around the centre of pdf, which is a good description only for certain businesses (credit institutions, high-yield funds, etc.).
[37] In practice, for many different reasons, multiple-periods reinvestment with stable quantifiable edge is rarely possible. The volatility (noise), the uncertainty and the competitive deterioration of any edge are very powerful drivers of the markets, which make the sustainability of the edge very difficult and very rare.

Taking advantage of the diversity breakdowns (M. Mauboussin 2007) (regime changes) and benefiting from the expected value mindset (M. Mauboussin 2004) take us into the heart of the realistic and successful management of grey swans. "Dramatic diversity breakdown" and "time arbitrage breakdown" are important market heuristics of the regime changes, which are characteristic of the complex adaptive systems. Diversity breakdowns become triggered as more and more investors recognize particular patterns (viewed as trends or risk premiums), initiated at different times and managed to different time horizons, and deploy higher leverage under collectively perceived and diminishing uncertainty. Subjective views become less diverse resulting in the diversity breakdowns. Risk premium strategies, which can be levered according to Kelly criterion, appear to compete with the diversity breakdowns, which spark sharp or even violent trends. In any case, open competition of the investors' strategies in the allocation of capital makes good markets. Another aspect is that liquidity has to be commensurate with the leverage and the investor strategy. Liquidity is a very important part of the feedback loops, where the investor provides and depends on the liquidity from the complex markets. In practice, risk taking not only can be empowered by the statistical quantitative methods (like Kelly criterion) but also has to evolve by adapting to the regime changes and the diversity and liquidity breakdowns.

Uncertainty and leverage are the key risk preferences. Prices (information and patterns), preferences (aversions, uncertainty, and leverage), and probabilities (risk) have been highlighted to encompass consistent risk management (Lo 1999). Risk taking and management are about consilience of qualitative, quantitative, technical and behavioural skills, which come together, in particular, in the judgement about the uncertainty and choosing the leverage accordingly. Risk management and risk taking are dual. The same events, the same grey swans, bring dangers and opportunities associated with uncertainty. Interestingly, on general grounds, in the entrepreneurial context (Knight 1964) thought that the uncertainty deserves much higher returns than the risk. Consistent and skilful management of the uncertainty and the leverage can be the source of structural edge, especially since the market risk premiums can be largely due to the uncertainty (Lemperiere 2015).

---

[38] Realistic markets also exhibit various memory and correlation effects in the risk and uncertainty pricing, and there is good value for risk takers in recognizing even the short-range correlations in multiplicative processes (see nice mathematical illustrations in (Redner 1990)). The pattern recognition is one of the salient properties of good risk takers; see for example, a classic book (Lefevre 2005).

# Conclusion

The leverage and the uncertainty have to be managed consistently in a proper, even though subjective, risk-management framework, and several practical results and guidelines have been discussed here. In particular, the uncertainty determines how fractional Kelly criterion can be used more effectively in practice. The fractional weight can be estimated, not exactly calculated, from the subjective probability weights and the severity of the uncertainty tails. Uncertainty, due to many uncontrollable and unpredictable events, constraints leverage beyond risk.

Kelly Parity, multi-asset portfolio based on Kelly criterion of optimisation of geometric growth rates, encompasses major, widely known, portfolio approaches of Risk Parity and Mean-Variance optimization. In particular, Kelly Parity becomes Mean-Variance Optimization if the absolute leverage is constrained to unity. Kelly Parity becomes basic Risk Parity if the Sharpe ratio of all involved assets is the same. Due to the volatility convexity and skew (parametrized by left and right tails), Kelly Parity has to adjust absolute and relative leverages.

"The fundamental law of investing is the uncertainty of the future" (Bernstein 1992). The management of leverage and uncertainty is one of the most important strands in successful investing. The paper presents a certain perspective and a programme for the future empirical and theoretical research with many open questions. Human decision-making, adaptiveness, and the drawdown management are discussed phenomenologically from the practitioner's perspective. Adaptive and skilful management of uncertainty and leverage can build a successful edge by taking on "grey swans" opportunities and managing possible drawdowns.

# Appendix A. Analytical calculation for the pdf with the tail loss probability.

The original Kelly criterion can be reproduced by using a pdf as a sum of the two delta functions. The pdf is of the form $P(x) = 0.5 \left(\delta(\mu + \sigma) + \delta(\mu - \sigma)\right)$, where the excess return $\mu$ and the volatility $\sigma$ are parameters. We assume the limit $\sigma \gg \mu$ for the consistency with Kelly calculations. The geometric growth rate (Thorp 2006) as a function of constant fraction of the invested capital $f$ is

$$g_0(f) = 0.5 \ln(1 + f(\mu + \sigma)) + 0.5 \ln(1 + f(\mu - \sigma)).$$

The maximization solution gives $f_0^* = \frac{\mu}{\sigma^2 - \mu^2} \approx \frac{\mu}{\sigma^2}$ and gives $g_0(f_0^*) \approx 0.5 \frac{\mu^2}{\sigma^2}$ consistently with the Thorp's results.

*A physical interpretation of Kelly leverage is the ratio of the diffusion time $1/\sigma^2$ divided by the drift time $1/\mu$.* In this interpretation, small allocation $f^* \ll 1$ means that the drift time $1/\mu$ is longer than the diffusion time $1/\sigma^2$, and the "return edge" is small (the geometric growth $g(f^*)$ is smaller than the return $\mu$). High allocation $f^* \gg 1$ means that the growth is fast enough, not eroded too quickly by the noise, in this case the drift time is shorter than the diffusion time.

We would like to analyze the results in the case of possible, but rare, large tail losses. The pdf can be defined as $P(y) = \alpha\, \delta(y + ETL) + \frac{1-\alpha}{2}[\delta(y - (\mu + \sigma)) + \delta(y - (\mu - \sigma))]$, where the additional parameters are the probability of the tail loss $\alpha$ and the size of the tail loss $ETL$ (Expected Tail Loss). For the same "positive edge", the condition $\mu > \alpha \cdot ETL$ is required. The relation $\frac{\mu_0}{\sigma_0^2} < \frac{1}{ETL}$ is required for the growth rate $g(f)$ staying well defined at the tail loss state. The regime $\frac{1}{ETL} < \frac{1}{\sigma}$ implies the tail loss is noticeably larger than the volatility.

The geometric growth function, to be maximized, is

$$g(f) = \alpha \ln(1 - f \cdot ETL) + \frac{1-\alpha}{2}[\ln(1 + f(\mu + \sigma)) + \ln(1 + f(\mu - \sigma))].$$

The relevant solution[39] continuously connected with the original $f^*$ is[40]

$$f_1^* \approx \frac{\mu - \alpha(\mu + ETL)}{\sigma^2 - \mu^2 + \alpha(2\mu ETL - (\sigma^2 - \mu^2))} \approx \frac{\mu}{\sigma^2} \frac{(1 - \frac{\alpha ETL}{\mu(1-\alpha)})}{(1 + \frac{\alpha ETL^2}{\sigma^2(1-\alpha)})}$$

$f_1^*$ is dependent on the tail loss $ETL$ in the two combinations $\frac{\alpha ETL}{\mu(1-\alpha)}$ and $\frac{\alpha ETL^2}{\sigma^2(1-\alpha)}$. Clearly, $f_1^* < f^*$ is reduced by numerical factor, and the growth rate is smaller as well

$$g(f_1^*) \approx \frac{\mu^2}{2\sigma^2}(1-\alpha)\frac{\left(1 - \frac{\alpha ETL}{\mu(1-\alpha)}\right)^2}{(1 + \frac{\alpha ETL^2}{\sigma^2(1-\alpha)})}$$

---

[39] The model considered here is a static model based on the pdf. It would be interesting to generalize the model to the dynamic model and the limit of large N-steps (p.22 of (Thorp 2006)). The time scale related to the tail loss mixing and the scaling with N would be of interest.

[40] Perhaps, some of these results are due to the particular form of the logarithm dependence. Notice that the exponential growth function is naturally logarithmic. The growth function is conceptually different from the conventional utility function of the probabilistic utility theory.

It is interesting that the effect of the relative impact parameter $\left(1 - \frac{\alpha ETL}{\mu(1-\alpha)}\right)$ on the growth rate $g(f_1^*)$ is only quadratic. We can see clearly[41] that the growth rate decreases sharply when $2\alpha ETL$ becomes of the same order as the excess return $\mu$, this condition is more stringent by a factor of 2 than the threshold for the positivity of the arithmetic growth rate (also the first moment) $-\alpha \cdot ETL + (1-\alpha)\mu > 0$. This result illustrates the robustness of the Kelly system to the perturbative tail effects, but the adjustment of the leverage is esssential nevertheless. A dynamical model would shed more light on the different timescales associated with the drift, diffusion, the tail loss and their mixing.

The above analytical results were tested by the exact numerical solution of the lengthy quadratic equation for $f_1^*$. The parameters of the return and the volatility can be recalibrated to the original parameters as follows. The mean and the variance are given by the expresions $<y> = -\alpha \cdot ETL + (1-\alpha)\mu$ and $<y^2> = \alpha \cdot ETL^2 + (1-\alpha)(\mu^2 + \sigma^2)$. If the original excess return $\mu_0$ and the volatility $\sigma_0$ are substituted, then the calibrated values of $\mu$ and $\sigma$ are $\mu = \frac{\mu_0}{1-\alpha} + \frac{\alpha}{1-\alpha}ETL$ and $\sigma^2 = \frac{\sigma_0^2}{1-\alpha} - \alpha(ETL + \mu)^2$. One observation, which emerges from the numerical parameter testing, is that the the growth rate can remain stable. At the same time, the leverage needs to be increased significantly when $ETL$ increases. A qualitative interpretation of this numerical finding is that the tail loss can remain hidden and unnoticed at the level of average growth. But many path-dependent properties do change due to the tail loss (for instance, the drawdown distribution), especially in the view of slow convergence of the multiplicative process.

Another observation is that for sufficiently strong tail loss $ETL$ the solution $f_1^*$ switches from positive to negative smoothly. This means that for strongly skewed asset it is optimal to "short the asset". This result is shown in the Figure below.

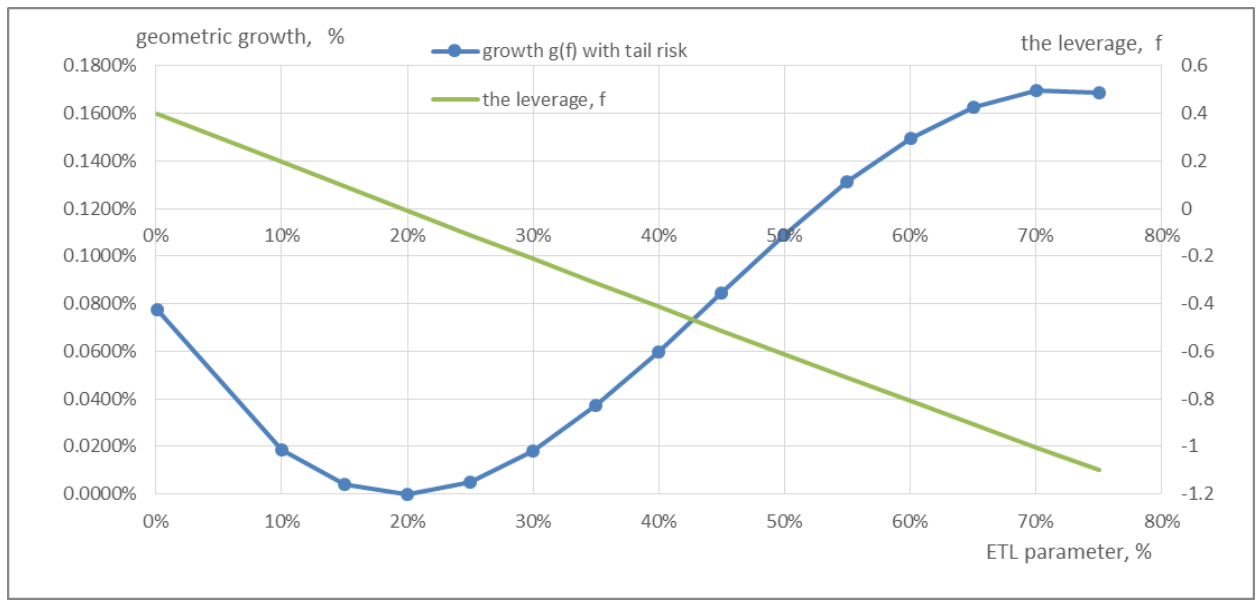

---

[41] The full set of the conditions is (a) $\sigma \gg \mu$ for the consistency of the low leverage approximation of the assumed pdf (b) $\frac{\mu_0}{\sigma_0^2} < \frac{1}{ETL}$ to preserve the original maximum of the growth function (c) $\alpha \ll 1$ for the perturbative effect of the tail loss (d) $\frac{ETL}{\mu} > 1$ for the meaningful parameter choice.

The geometric growth and the leverage are shown as a function of $ETL$. The parameters are $\mu = 0.4\%, \sigma = 10\%, \alpha = 2\%$. Simple Kelly leverage and growth rate are $f_0^* = 0.4$ and $g_0^* = 0.08\%$. Notice that the arithmetic growth is negative $r = \mu - \frac{\sigma^2}{2} = -0.1\%$. For $ETL = 10\%$, the fractional leverage is $f_1^* = 0.2$, and the fractional growth rate drops to $g_1^* = 0.02\%$.

In a separate exercise, in a similar way how the volatility is described to be convex and skewed, we can ask what the skew and the convexity of the growth rate are. The leverage and the geometric growth decrease as the volatility increases, which mean that Kelly strategy has a negative skew as a function of the volatility. The second derivative $d^2 g^*/d^2\sigma$ is positive, thus the Kelly strategy has a positive convexity as function of the volatility. The volatility of a stock (to be specific) has the skew and the convexity as a function of the stock price. Thus we can express the non-linear properties of the geometric growth rate as a function of the stock price[42]. Straightforwardly, the growth skew $Skew_g$ and the growth convexity $Convexity_g$ are scaled by the same factor $Z_g = -dg^*/d\sigma + 2\sigma d^2g^*/d^2\sigma$ from the corresponding volatility skew $Skew_\sigma$ and the volatility convexity $Convexity_\sigma$. Since the scaling factor $Z_g = 13\mu/\sigma^3$ for simple Kelly case is positive for the above form of $g^*(\sigma)$, the growth rate has the same sign skew $Skew_g = Z_g \cdot Skew_\sigma$ (negative for equities) and the same sign convexity $Convexity_g = Z_g \cdot Convexity_\sigma$ (positive for equities) as a function of the stock price.

## Appendix B. Concave Efficient Frontier

The main quantitative thrust of the paper focuses on the optimization of the geometric growth rate by extending Kelly criterion in the presence of the fat tails for the single asset and multi-asset portfolio. The larger qualitative problem, addressed in the paper, is the critical role of the leverage and the uncertainty in the trading and investment decisions, which are not addressed by the classical theories of mean-variance optimisation and CAPM (Capital Asset Pricing Model). Although the central role of the leverage and fat tails (the uncertainty) is a completely different starting point fundamentally and conceptually from those well-established theories, nevertheless it is instructional (and the topic of this appendix) to discuss some of the perturbative changes to the classical theories due to the the leverage and the uncertainty. These insights allow comparing and relating different theoretical vantage points.

The investors are very concerned with possible large losses or drawdowns. Among various important risk preferences of the investors is the drawdown loss aversion, which can be defined as the degree of the drawdown (the loss of the capital) acceptable for the investor. In other words, let us imagine that the investor specifies the acceptable drawdown level (the loss cannot exceed this level) and accepts the consequences of the choice[43]. The drawdown history and analysis are already a noticeable practice in the financial industry. The drawdown aversion is one of the quantitative ways to quantify the loss aversion and estimate the subjective uncertainty. The perceived uncertainty and the loss aversion determine subjective investor's drawdown level.

To model the drawdown aversion, the investor has to buy the downside put option with the strike at the drawdown level. The cost of this option modifies the efficient frontier. The straight-line efficient frontier becomes *strongly concave frontier, which is due to the skew* of the probability distribution (asymmetry between losses and gains) and *due to convexity* (heavy fat tails without asymmetry between tails). Naturally, the efficient frontier becomes lower. Even without the skew and smile effect on the volatility, the put option cost lowers the return. The skew and smile volatility effects make the tails of the pdf heavy (fat), which make in turn the put option cost more significant and decrease the expected return further.

---

[42] A simple static, so called local, volatility skew and convexity are assumed.
[43] This risk preference will require the investor to understand the costs of the option protection. The investor will have to weigh the comfort of the drawdown protection and the reduced performance target (due to the cost of the protection).

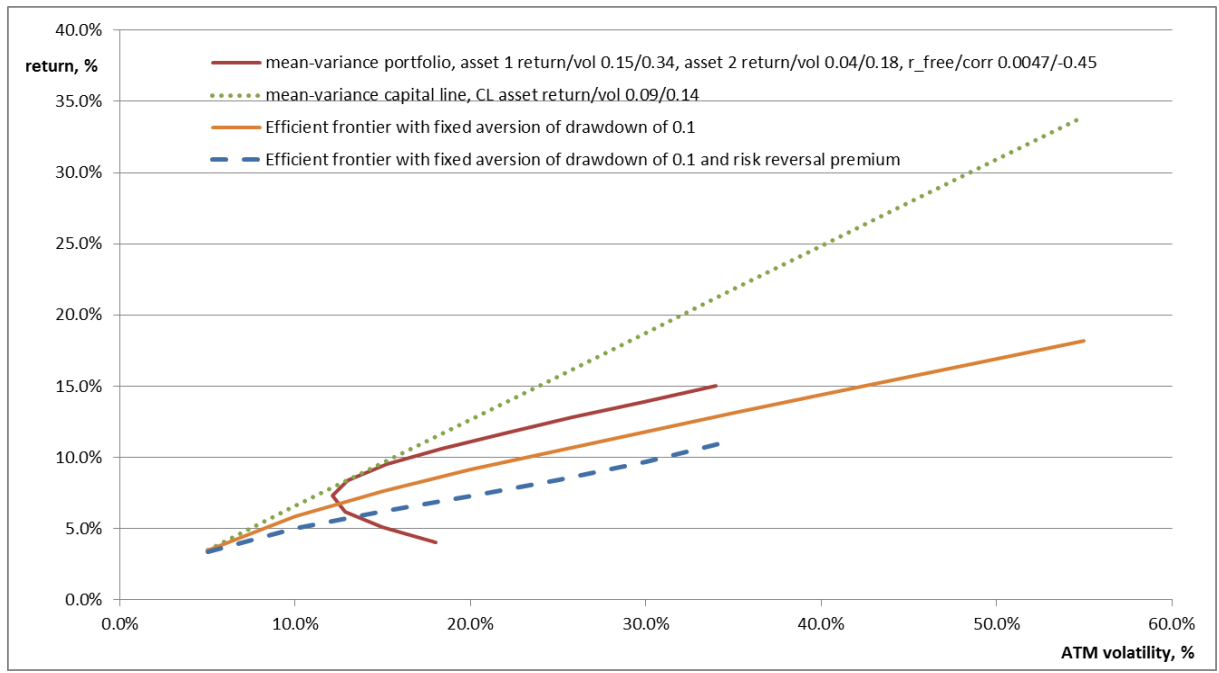

Fig 2. The efficient frontier shifts lower because of the drawdown protection cost (acceptable 10% loss) without and with additional skew (risk reversal premium) affect.

Another important point is that the efficient frontier is not unique anymore for all investors, because the level of drawdown aversion differentiates different investors in their expected returns (drawn as a function of the volatility, the conventionally defined risk). Although in practical MPT (modern portfolio theory) application, investors may differ in their projected expected returns and risks, the differentiation between investors due to their drawdown aversions may be expected to be even more significant[44].

In addition to the drawdown aversion, the mechanisms of the leverage financing also add to the concavity of the efficient frontier. Stronger concavity, for the part of the efficient frontier with the leverage above unity, can be related to the margin financing spread, paid by the leverage taker, and various so-called wrong way risks related to gap risks and outflows risks (for instance, (M. J. Mauboussin 2006)). The latter risks lead to reduced returns because the leverage unwinds happen "at the worst times" or "at the least favourable terms". The concave shape and even downward sloping of the efficient frontier, due to stop-loss and overleveraging, were suggested before by (Illinski 2010).

The importance of the drawdown aversion can be highlighted in another way. Different strategies and portfolios can aim to achieve high returns by taking significant downside risks. In order to put different funds (strategies) on the same footing and prevent over-leveraging of the downside risks, DrawDown Valuation Adjustment (DDVA), which quantifies the cost of the drawdown protection for a given drawdown level, can be introduced. This type of the valuation adjustment is similar in spirit to the so called CVA (Credit Valuation Adjustment) from the derivatives industry (Gregory 2012). CVA expresses a valuation adjustment to MtM (Mark-to-Market) value of the derivatives due to the counterparty credit risk not mitigated by any collateral. CVA eliminated the arbitrage between the cash instruments (bonds, etc.) and the derivative instruments. Similarly, the Drawdown VA should eliminate the arbitrage between the funds taking excessive risks in the sense of possible significant loss and the funds taking more balanced risks. The conceptual difference between CVA and DDVA is that Drawdown VA is proposed to be calculated below a certain drawdown strike, corresponding to a certain level of loss. Thus a certain range of the volatility is not penalized. DDVA at "certain loss level" is consistent with the main risk being "a permanent loss", where the investor cannot accept further losses beyond the drawdown level, while she accepts the volatility at the "normal course of events". Such an adjustment would be challenging the financial industry to implement for many reasons, from technical to educational; nevertheless, this adjustment reflects a realistic value adjustment due to loss uncertainties. In the context of MPT, Markowitz stressed the applicability of the utility theory beyond Gaussian distributions, but he did not consider explicitly the drawdown aversion and multi-period growth rate calculated by the Kelly criterion.

---
[44] The heterogeneity of the investors is appropriate in the description of the markets.

In summary, by introducing DDVA adjustment due to the drawdown aversion within the framework of MPT, we noticed that the risk premium on the efficient frontier (per unit of asset) depends not only on the volatility (as in CAPM) but also non-linearly on the drawdown aversion and the leverage. This conclusion highlights that there are non-linear feedback effects between the investor's choices of the leverage, the views on the uncertainty and the expected returns.